\documentclass[conference]{IEEEtran}
\IEEEoverridecommandlockouts
\usepackage{cite}
\usepackage{amsmath,amssymb,amsfonts}
\usepackage{graphicx}
\usepackage{textcomp}
\usepackage{xcolor}
\usepackage{enumitem}
\usepackage{stfloats}
\usepackage[all=normal,charwidths]{savetrees}

\def\BibTeX{{\rm B\kern-.05em{\sc i\kern-.025em b}\kern-.08em
    T\kern-.1667em\lower.7ex\hbox{E}\kern-.125emX}}

\usepackage{comment}

\usepackage{algorithm}
\usepackage[noend]{algpseudocode}

\usepackage[T1]{fontenc}
\usepackage[utf8]{inputenc}
\usepackage{soulutf8}
\usepackage{color}

\usepackage{array}
\usepackage{booktabs} 
\usepackage{multirow}

\usepackage{numprint}
\usepackage{nameref}
\usepackage{url}
\usepackage[hidelinks]{hyperref}
\usepackage{amsmath}
\usepackage{amsfonts}
\usepackage{amssymb}
\usepackage{xcolor}

\usepackage[T1]{fontenc}
\usepackage{colortbl}

\usepackage{babel}
\definecolor{Gray}{gray}{0.85}
\definecolor{LightGray}{gray}{0.95}
\definecolor{LightCyan}{rgb}{0.88,1,1}

\usepackage{color}
\definecolor{comments}{rgb}{0.13,0.55,0.13}
\definecolor{background}{rgb}{0.94, 0.97, 1.0}
\definecolor{strings}{rgb}{0.63,0.125,0.94}
\usepackage{listings}

\usepackage{tikz}
\usepackage{scalerel}
\usepackage[most]{tcolorbox}
\usepackage{setspace}
\usepackage{indentfirst}

\newcommand{\tikztriangleright}[1][red,fill=red]{\scalerel*{\tikz \draw[rounded corners=0.1pt,#1] (0,-2.5pt)--++(0,5pt)--++(-30:5pt)--cycle;}{\triangleright}}

\tcbset{
    enhanced,
    breakable,
    left=1mm,
    right=1mm,
    attach boxed title to top left={
        xshift=0.5cm,
        yshift= -3.5mm
    },
    top=3mm,
    bottom=0pt,
    colframe=blue!75!black,
    coltitle=white,
    colbacktitle=black!70!white,
    fonttitle=\footnotesize\bfseries,
    fontupper=\scriptsize\sffamily,
}

\newenvironment{mybox}[1]{%
    \begin{tcolorbox}[title={#1}]%
    \setstretch{0.95}}{
    \end{tcolorbox}
}

\makeatletter
\def\lst@makecaption{%
  \def\@captype{table}%
  \@makecaption
}

\usepackage{tikz}

\newcommand{\framework}[1]{\textit{RunPHI}}

\newcolumntype{L}[1]{>{\raggedright\arraybackslash}m{#1}}
\newcolumntype{C}[1]{>{\centering\arraybackslash}m{#1}}
\newcolumntype{R}[1]{>{\raggedleft\arraybackslash}m{#1}}

\begin{document}


\title{{RunPHI}: Enabling Mixed-criticality Containers via Partitioning Hypervisors in Industry 4.0}


\author{\IEEEauthorblockN{Marco Barletta, Marcello Cinque, Luigi De Simone, Raffaele Della Corte, Giorgio Farina, Daniele Ottaviano}
\IEEEauthorblockA{\textit{DIETI - Università degli Studi di Napoli Federico II, Via Claudio 21, 80125 Napoli, Italy}\\
\{marco.barletta, macinque, luigi.desimone, raffaele.dellacorte2, giorgio.farina, daniele.ottaviano\}@unina.it}
}

\maketitle


\thispagestyle{plain}
\pagestyle{plain}

\begin{abstract}
Orchestration systems are becoming a key component to automatically manage distributed computing resources in many fields with criticality requirements like Industry 4.0 (I4.0). However, they are mainly linked to OS-level virtualization, which is known to suffer from reduced isolation. In this paper, we propose \framework{} with the aim of integrating partitioning hypervisors, as a solution for assuring strong isolation, with OS-level orchestration systems. The purpose is to enable container orchestration in mixed-criticality systems with isolation requirements through partitioned containers.
\end{abstract} 





\section{Introduction}
Nowadays, we are witnessing the spread of Information Technologies in several industrial domains (e.g., railways, avionic, automotive). This transforms industrial scenarios in Edge cloud environments populated by many Industrial Internet of Things (IIoT), looking towards the \textit{Industry 4.0} (I4.0) vision \cite{iiot_survey, stavdas2022networked}. 
Thus, these systems must meet not only mandatory regulatory requirements involving functional safety and control timeliness, but also performance scalability, interoperability, low latency, and reconfigurability through fast and efficient deployment.

Virtualization is an enabling technology for I4.0 since it responds to the needs of reconfiguration, modularity, and consolidation through resource partitioning and multiplexing. It allows the execution of heterogeneous Operating Systems (OS) (real-time and general-purpose) on the same system-on-a-chip (SoC), becoming a prominent way for the industry to realize Mixed-Criticality systems due to the current COVID-19-induced silicon shortage phenomenon \cite{bloomberg_chip_shortage, cinque2021virtualizing, cilardo2021virtualization}. 
\textit{Partitioning hypervisors} (e.g., \textit{Jailhouse} \cite{ramsauer2017look}, \textit{Bao} \cite{martins2020bao}, \textit{Xtratum} \cite{crespo2010partitioned}) have gained the attention of both academia and industry \cite{hermes_project, selene} due to the strong isolation provided through static allocation, at the cost of a reduced flexibility of deployment compared to classical virtualization (recently named \textit{consolidating hypervisors}).

However, in the I4.0 vision, the stress is on the automatic management, reconfiguration, and self-healing of IT systems. Thus, \textit{criticality-aware orchestration systems} are paramount since they automatically place, deploy, monitor, and migrate the packaged software across the infrastructure
\cite{barletta2022achieving}; still being aware of the isolation guarantees required by critical workloads to prevent interferences in terms of faults and attacks from non-critical jobs. 
Currently, containers are seamlessly integrated into orchestration systems, but ensuring their isolation is still an open issue, threatening the practicability of OS-level virtualization under strict real-time, safety, and security requirements. Unikernels seem to be a solution since they do not share the underlying host kernel, but their portability issues and real-time support are still open issues \cite{chen2022unikernel}.

In this position paper, we propose \framework{}, a framework that integrates partitioning hypervisors into container orchestration systems with the aim of leveraging the strong isolation provided by partitioning solutions while taking advantage of orchestration techniques. This project advances the state of the art since \textit{i)} it simplifies the deployment of critical workloads in edge/cloud environments, useful for maintenance, upgrades, and new deployments; \textit{ii)} it enables failure mitigation through migration and spawning of new partitions; \textit{iii)} it is a driving force for the full reconfigurability of I4.0 for workloads with isolation requirements.

\vspace{-0.2cm}
\section{Related Work}

In the literature, several solutions (summarized in \tableautorefname{}~\ref{tab:table1}) adapt general-purpose hypervisors with the aim of providing true isolation between containers, aka \textit{sandboxed containers}. \textit{IBM Nabla}\footnote{https://nabla-containers.github.io/} builds containers on top of unikernels. \textit{Google gVisor}\footnote{https://gvisor.dev/} creates a dedicated guest kernel to run containers. \textit{Amazon Firecracker}\footnote{https://firecracker-microvm.github.io/} is a lightweight hypervisor for sandbox applications. Both \textit{KubeVirt}\footnote{https://kubevirt.io/} and \textit{vSphere Integrated Containers (VIC)}\footnote{https://vmware.github.io/vic-product/} integrate VMs and containers under a single orchestration infrastructure. \textit{Kata Containers}\footnote{https://katacontainers.io/} allows running secure container runtime with lightweight VMs. \textit{RunX}\footnote{https://github.com/Xilinx/runx} uses Xen hypervisor to run containers in multiple separate VMs, either with the provided custom-built Linux-based kernel, or with container-specific kernel/ramdisk.

\vspace{-0.5cm}
\begin{table}[ht]
\centering
\caption{State-of-the-art solutions for partitioned containers.}
\label{tab:table1}
\footnotesize                       
\resizebox{\linewidth}{!}{%
\sffamily \setstretch{0.50}
\begin{tabular}{C{3cm}C{2.3cm}C{2.5cm}C{3cm}} \hline 

\rowcolor{LightCyan}
 \textsf{\textbf{Solution}} & 
 \textsf{\textbf{Guest Type}} & 
 \textsf{\textbf{Used Hypervisor}} &
 \textsf{\textbf{Orchestration Support}}\\ \hline \hline 

\textsf{\textbf{Nabla Container}} &
\textsf{Unikernel} & \textsf{\textit{Nabla Tender}} & \textsf{ Docker} \\ \hline 

\rowcolor{LightGray}\textsf{\textbf{gVisor}} & 
\textsf{Container + user-space kernel} & \textsf{KVM} & \textsf{ Kubernetes, Docker}\\ \hline 

\textsf{\textbf{Firecracker}} &
\textsf{Light VM} & \textsf{KVM} & \textsf{OCI compliant}\\ \hline 
\rowcolor{LightGray}\textsf{\textbf{KubeVirt}} & \textsf{VMs and Containers} & \textsf{KVM} & \textsf{ Kubernetes}\\ \hline 

\textsf{\textbf{vSphere Integrated Containers (VIC)}} & \textsf{VMs and Containers} & \textsf{VMware ESXi} & \textsf{VMware Orchestrator, Docker}\\ \hline 

\rowcolor{LightGray}\textsf{\textbf{KataContainer}} &
\textsf{Light VM} & \textsf{QEMU/KVM} & \textsf{ Kubernetes, Docker}\\ \hline 

\textsf{\textbf{RunX}} & 
\textsf{Light VM} & \textsf{Xen} & \textsf{Kubernetes, Docker}\\ \hline 

\rowcolor{cyan}\textsf{\textbf{RunPHI}} & 
\textsf{Light VM} & \textsf{Partitioning Hypervisors} & \textsf{Kubernetes, Docker, OCI compliant} \\ \hline
\end{tabular}
}
\end{table}

These solutions are mainly based on general-purpose hypervisors, which do not fit well with mixed-criticality real-time requirements. In contrast, current partitioning hypervisor solutions seem to provide enough guarantees about both safety and security isolation. To the best of our knowledge, there are no solutions that support sandboxed containers in conjunction with partitioning hypervisors. The objective of \framework{} is to provide both isolation requirements and flexible orchestration capabilities for next-generation I4.0 scenarios.

\section{Proposal}





\begin{figure}
    \centering
            \includegraphics[width=\columnwidth]{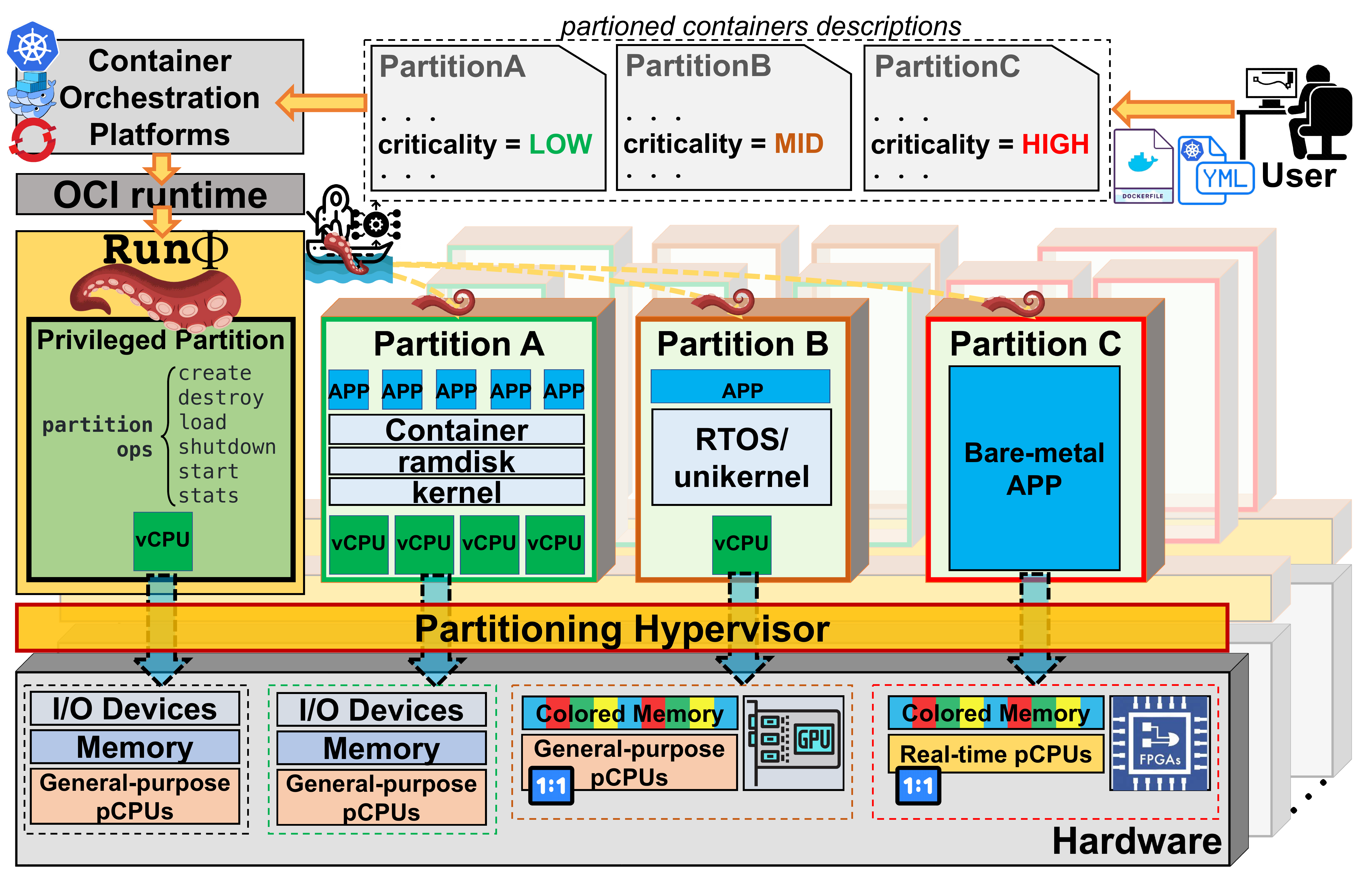}
    \caption{Proposed \framework{} Architecture.} 
    \label{fig:runj_architecture}
    \vspace{-.5cm}
\end{figure}

\figureautorefname{}~\ref{fig:runj_architecture} shows a first design of \framework{}. Users provide \textit{partition descriptions} via classical tools inherited from container-based orchestration (e.g., Dockerfiles, Kubernetes manifests). \textit{Partitioned container} descriptions can be extended with requirements related to physical resources, criticality levels (e.g., low, mid, or high), real-time constraints, etc. \framework{} leverages a partitioning hypervisor to provide strong isolation between containers. In particular, according to partition descriptions, \framework{}, implemented in the privileged partition, tries to allocate physical resources in line with free resources within the host node. 
The \textit{inference engine} fills the gaps with predicted values for resources not specified in the \textit{partitioned container} description, according to requests from the orchestration platforms and current usage of hardware. \framework{} manages the lifecycle of partitioned containers and is
designed to be highly flexible with the aim of orchestrating partitioned containers with a: i) \textit{low-level criticality} (e.g., partition A in \figureautorefname{}~\ref{fig:runj_architecture}) that includes a classical container abstraction with several applications running on top of it, a number of virtual CPUs (vCPUs) with no specific affinity on physical CPUs (pCPUs), without specific real-time guarantees; ii) \textit{mid-level criticality} (e.g., partition B in \figureautorefname{}~\ref{fig:runj_architecture}) that includes running a RTOS/unikernel single-app with strict temporal and memory isolation requirements (e.g., by using 1-to-1 vCPU-pCPU mapping and cache/RAM coloring mechanisms respectively) and use of GPUs accelerators for running machine learning algorithms; iii) \textit{high-level criticality} (e.g., partition C in \figureautorefname{}~\ref{fig:runj_architecture}) that includes same mechanism for mid-level criticality with the addition of running bare-metal tasks on real-time CPUs and use of programmable logic blocks like FPGAs.

\section{Research Questions and Objectives}
In the following, we delineate research questions to be considered in the next steps of our project.


    
        
        
        






    \begin{mybox}{\parbox{6cm}{\textit{RQ1.} How I4.0 mixed-criticality systems can be deployed via \framework{}?}}
    
        \textbf{Objectives:}
        
        $\tikztriangleright[blue,fill=gray!80]$ To support partitioned containers run at different criticality with real-time constraints
    
        $\tikztriangleright[blue,fill=gray!80]$ To support running bare-metal applications
        
        $\tikztriangleright[blue,fill=gray!80]$ To support accelerator devices like FPGAs and GPUs
        
        $\tikztriangleright[blue,fill=gray!80]$ To induce minimal overhead in terms of CPU, memory, and I/O
        
        
        
        
        
    \end{mybox}

    \begin{mybox}{\parbox{6cm}{\textit{RQ2.} How to quantify the isolation between partitioned containers provided by \framework{}?}}
    
        \textbf{Objectives:}
        
        $\tikztriangleright[blue,fill=gray!80]$ To support temporal, memory, and fault isolation assessment (e.g., fault injection testing) 
        
        $\tikztriangleright[blue,fill=gray!80]$ To support security isolation assessment (e.g., fuzzing)
        
    \end{mybox}
    
        
        
        
%


    \begin{mybox}{\parbox{6cm}{\textit{RQ3.} How to support orchestration for partitioned containers in \framework{}?}}
        
        \textbf{Objectives:}
        
        $\tikztriangleright[blue,fill=gray!80]$ To support different runtime containers and OCI-compliance
        
        $\tikztriangleright[blue,fill=gray!80]$ To support partitioned containers description via existing runtime containers API and existing tools for configuration file (e.g., Dockerfile, Kubernetes manifest)
       
        $\tikztriangleright[blue,fill=gray!80]$ To implement an inference engine to determine (sub)optimal resource allocation for partitioned containers 
        
        
        
        $\tikztriangleright[blue,fill=gray!80]$ To support migration, checkpointing, and high-availability mechanisms for partitioned containers

    \end{mybox}
    



\section*{Acknowledgment}
This work has been supported by the project COSMIC of UNINA DIETI.

\bibliographystyle{IEEEtran}
\bibliography{bibliography}

\end{document}